\documentclass{article}
\usepackage{ijcai/ijcai25}

\usepackage{times}
\usepackage{soul}
\usepackage{url}
\usepackage[hidelinks]{hyperref}
\usepackage[utf8]{inputenc}
\usepackage[small]{caption}
\usepackage{graphicx}
\usepackage{amsmath}
\usepackage{amsthm}
\usepackage{booktabs}
\usepackage{algorithm}
\usepackage{algorithmic}
\usepackage[switch]{lineno}
\usepackage{color}
\usepackage{amssymb}
\usepackage{multicol}
\usepackage{multirow}
\usepackage{subfigure}
\usepackage{enumitem}

\title{UFO: Unfair-to-Fair Evolving Mitigates Unfairness in LLM-based Recommender Systems via Self-Play Fine-tuning}

\author{
Jiaming Zhang$^1$
\and
Yuyuan Li$^2$
\and
Xiaohua Feng$^1$
\and
Zhifei Ren$^3$
\and
Li Zhang$^1$
\And
Chaochao Chen$^1$\thanks{Corresponding author}
\affiliations{
$^1$Zhejiang University\\
$^2$Hangzhou Dianzi University\\
$^3$Southeast University
}
\emails{
jm.zhang@zju.edu.cn,
y2li@hdu.edu.cn,
fengxiaohua@zju.edu.cn,
213231895@seu.edu.cn,
zhanglizl80@gmail.com,
zjuccc@zju.edu.cn
}
}

\begin{document}

\maketitle

\newcommand{\model}{\texttt{UFO}}
\newcommand{\zjm}[1]{\textcolor{blue}{[#1--zjm]}}
\newcommand{\fxh}[1]{\textcolor{blue}{[#1--fxh]}}
\newcommand{\lyy}[1]{\textcolor{red}{[#1 --lyy]}}

\begin{abstract}
    Large language model-based Recommender Systems (LRSs) have demonstrated superior recommendation performance by integrating pre-training with Supervised Fine-Tuning (SFT).
    However, this approach introduces item-side unfairness. Existing studies primarily attribute this issue to the absence of fairness constraints during SFT and attempt to mitigate unfairness via re-weighting and re-ranking methods.
    In this paper, we find that unfairness arises not only from SFT but also from pre-training, where inherent biases are further amplified during SFT. This finding underscores the failure of current methods to address the root causes of unfairness. Moreover, current methods struggle to preserve satisfactory recommendation performance.
    To tackle these issues, we propose an Unfair-to-Fair evOlving (\model{}) framework using a self-play mechanism, formulating unfairness mitigation as a two-player game. \model{} alternates between two player roles: the \textit{judger}, which identifies unfairness from both pre-training and SFT, and the \textit{corrector}, which adjusts the LRS to address identified unfairness while preserving recommendation performance. Iterative optimization between these roles enables \model{} to completely resolve unfairness.
    Extensive experiments demonstrate that \model{} effectively mitigates unfairness while improving recommendation performance.
\end{abstract}

\section{Introduction}
Recommender Systems (RSs) act as personalized digital curators, optimizing user experiences by providing tailored content such as news~\cite{turcotte2015news}, videos~\cite{zhou2010impact}, medications~\cite{bao2016intelligent}, and jobs~\cite{paparrizos2011machine}, thus improving daily life and simplifying decision-making.
However, traditional RS are constrained to relying on historical interaction data from users and are limited to fixed input-output base models. 
% However, traditional RSs rely solely on historical interaction data and fixed input-output models.
%
Due to the absence of general world knowledge and the inability to dynamically integrate user needs, traditional RS are limited in their ability to capture complex user preferences and provide personalized recommendations. 
% Lacking general world knowledge and dynamic integration of user needs, they struggle to capture complex preferences and deliver truly personalized recommendations.

Recently, integrating Large Language Models (LLMs)~\cite{achiam2023gpt,touvron2023llama} with RSs has emerged as a popular approach to enhance the recommendation performance and personalization of recommendations~\cite{xi2024towards,zhang2023recommendation}. 
LLMs employed as foundational architectures or auxiliary tools to enhance RS, with their extensive knowledge base and robust instruction-following capabilities.
This integration typically involves two stages: pre-training and Supervised Fine-Tuning (SFT), a standard paradigm for applying LLMs to downstream tasks.

Although LLM-based Recommender Systems (LRSs) have yielded substantial benefits, it also raises pressing issues, particularly in the domain of fairness for both users and items~\cite{zhang2023chatgpt,tommasel2024fairness,jiang2024item}. 
Given that LRSs are frequently applied to sequential recommendation tasks, Item-side Fairness (IF) emerges as a primary concern.
IF refers to the principle that different item groups should be treated as equitably as possible~\cite{li2023fairness,wang2023survey}. Note that LRSs exhibits more severe unfairness than models like SASRec~\cite{kang2018self}, a representative of traditional sequential recommendation models, leading to severe inequities in item exposure and selection.
LRSs exhibit more severe unfairness compared to traditional recommendation models, resulting in significant inequality in item exposure and selection.
This has significantly undermined the interests of certain groups, such as job or news providers focused on specific topics.

Recent studies~\cite{jiang2024item} consider the incorporation of fairness constraints during the SFT stage as a primary approach to mitigate such unfair issue. Specifically, these studies simply employ re-weighting samples during the SFT stage and re-ranking the final recommendation lists based on fairness metrics. Although these methods alleviate unfairness to some extent, they still suffer from the following key issues:
\begin{itemize}[leftmargin=*]\setlength{\itemsep}{2pt}
    \item \textbf{Issue 1: insufficient analysis of unfairness causes.} Existing studies do not adequately analyze root causes of unfairness, particularly overlooking fairness issues during pre-training and the interplay between pre-training and SFT.
    \item \textbf{Issue 2: limited method applicability.} Current re-weighting and re-ranking methods are tailored to address unfairness in the SFT stage, lacking effectiveness for more complex, multifaceted unfairness challenges.
    \item \textbf{Issue 3: overlooking recommendation performance.} Existing methods fail to balance fairness enhancement with the recommendation performance of the LRS, lacking constraints that ensure high-quality recommendation results.
\end{itemize}

To address these issues, we systematically analyze the causes of unfairness in LRSs. Our analysis reveals that unfairness in the pre-training stage, which is further amplified during SFT, serves as the root cause. This explains why LRSs often exhibits worse fairness performance compared to traditional recommendation models. Building on this finding, we propose a framework called Unfair-to-Fair evOlving (\model{}) based on a self-play mechanism to mitigate the complex unfairness in LRSs effectively. \model{} iteratively identifies and addresses unfairness introduced during both pre-training and SFT, progressively optimizing the LRSs toward fairness. The framework operates as a two-player game between a \textit{judger} and a \textit{corrector}. The judger identifies unfair outputs from the current LRS and distinguishes them from ideal fair results, while the corrector adjusts recommendations to reduce unfairness, making it harder for the judger to classify results as biased. To enable the LRS to better operate at the distributional level, we introduce a simple distributional next-item generation mechanism, allowing the model to explicitly output an approximation of the recommendation distribution. In each iteration, the LRS is fine-tuned to generate recommendations aligning with a fair distribution, incrementally enhancing fairness. Finally, to preserve the recommendation performance of the original LRS, we introduce a geometric mixture policy. Specifically, the target LRS is updated via geometric mixture between the current and original models, ensuring that improvements in fairness are accompanied by corresponding gains in performance.

Our contributions are highlighted as follows:
\begin{itemize}[leftmargin=*]\setlength{\itemsep}{2pt}
    \item Regarding \textbf{issue 1}, we conduct a systematic analysis of unfairness in LRSs. 
    % Through SFT amplifies the unfairness of pre-training, the two stages are connected
    We find that unfairness exists in both the pre-training and SFT stages, with amplification occurring during SFT, providing a comprehensive explanation of the causes of unfairness.
    % explaining why LRS exhibits greater unfairness compared to traditional recommendation models.
    \item Regarding \textbf{issue 2}, we propose a self-play-based fairness correction method \model{}, a flexible post-training approach, which iteratively mitigates complex unfairness introduced by diverse sides through self-play fine-tuning. In different iterations, the LRS alternates between serving as the \textit{judger} and the \textit{corrector} within self-play game, iteratively identifying and mitigating unfairness.
    \item Regarding \textbf{issue 3}, we further incorporate a geometric mixture policy into our proposed framework, ensuring LRS does not deviate too far from the original one. This policy maintains recommendation performance while ensuring improvements in fairness.
    \item We conduct comprehensive experiments on multiple real-world datasets. For IF, \model{} achieves substantial fairness improvement while simultaneously enhancing recommendation utility. Additionally, we conduct extensive ablation studies on each module of our framework as well as key hyperparameters.
\end{itemize}
\section{Related Work}
\subsection{Recommender Systems based on LLMs}
Large Language Models (LLMs), with their vast world knowledge and advanced text generation and reasoning abilities, have recently gained prominence as both foundational models and auxiliary tools in recommender systems. 
The utilization of LLMs as a novel backbone for recommendations requires designing a pipeline optimized for recommendation tasks.
Early methods enhance recommendation generation by creating effective prompt templates or introducing additional information~\cite{lyu2023llm,sheng2024language}. 
To better align with the requirements of recommendation tasks, recent mainstream methods fine-tune LLMs on recommendation datasets~\cite{bao2025bi}. 
Some studies go further by modifying the model architecture and introducing special tokens~\cite{liao2024llara,qu2024tokenrec}. 
Preference Optimization~\cite{rafailov2024direct,meng2024simpo}, a common practice for LLMs after SFT, is also explored both online and offline in LRSs~\cite{bai2024aligning,liao2024rosepo}. 
For the purpose of enhancing the factual or personalization of LRSs, various works integrate techniques such as Retrieval-Augmented Generation~\cite{wang2025knowledge}, online feedback mechanisms~\cite{sun2024rlrf4rec}, and multi-agent collaboration~\cite{wang2024macrec} into the original pipeline.

While these approaches improve recommendation performance, they often build upon SFT for recommendation and exhibit significant issues in fairness. 
\vspace{-10pt}
\subsection{Item-side Fairness in Recommender Systems}
The fairness in RSs is typically categorized into individual fairness~\cite{dwork2012fairness,wang2022providing} and group fairness (the latter is based on attributes such as popularity or item category)~\cite{stratigi2020fair,wang2022make}.
Our research focuses on group fairness from the item side, advocating equal treatment across different groups. %defined by specific criteria.
In traditional RSs, achieving IF has primarily been approached in a staged manner. Relevant studies focus on i) IF-aware distribution adjustment of data before training~\cite{ekstrand2018all,rastegarpanah2019fighting}, ii) adjusting the loss function or introducing regularization terms during training~\cite{burke2018balanced,beutel2019fairness}, and iii) post-training adjustments such as modifying the ranking process~\cite{patro2022fair,zehlike2022fairness}. 

Notably, in LRSs, numerous studies~\cite{zhang2023chatgpt,deldjoo2024fairevalllm} have highlighted this unfairness issue, providing preliminary analyses of its causes. 
However, only a few have applied straightforward adaptations of %conventional 
methodologies from traditional recommendations~\cite{jiang2024item}, such as re-weighting during SFT or re-ranking unfair results. 
Additionally, D3~\cite{bao2024decoding} tackles item homogeneity during decoding, which can also be categorized as a post-processing adjustment to the ranking procedure. SPPO~\cite{gao2025sprec} leverages self-generated samples from the LRS as negative examples to reduce the bias of DPO. However, it remains grounded in the Bradley–Terry pairwise preference likelihood, which performs pointwise comparisons and thus struggles to align the overall distribution, offering limited global control over group-level fairness.

Existing approaches tend to superficially leverage statistical information from a certain dimension of the recommendation data to enhance IF through fairness-aware constraints. %
These approaches overlook the characteristics of LLMs and their complex causes of unfairness from both pre-training and SFT, which are not present in traditional recommendations. 
%
% Existing methods focus on a single unfair dimension (grouping attribute), addressing unfairness only during SFT and ignoring unfairness from pre-training. We propose a Self-Play  Fine-tuning approach that allows the LRS to autonomously detect and correct multidimensional unfairness, improving fairness across multiple dimensions.
\section{Preliminary}
\subsection{LRS for Sequential Recommendation}
Sequential recommendation can be formalized as a \textbf{next-item prediction} problem.  
Let the item space be \(\mathcal{I} = \{i_1, i_2, \dots, i_{|\mathcal{I}|}\}\).  
Each recommendation instance is represented by an item sequence \(s = \{i_1, i_2, \dots, i_{L}\}\), where \(L\) denotes the sequence length.  
Given a prefix sequence \(s\), the goal is to predict the next item from the item space:
\[
\hat{i} = \arg\max_{i \in \mathcal{I}} P(i \mid s).
\]

We denote the training dataset as \(\mathcal{S} = \{(s_j, i_j)\}_{j=1}^N\),  
where \(s_j\) is an input sequence and \(i_j\) is its corresponding target item. LRSs leverage LLMs to perform next-item prediction.  
Since LLMs are inherently trained under the \textbf{next-token prediction} paradigm,  
LRSs can naturally utilize LLMs for sequential recommendation by representing each item with its textual information (e.g., title, description).

The per-sample training loss is denoted by \(\ell(\boldsymbol{\theta}; s, i)\),  
which in LRSs typically corresponds to the supervised fine-tuning (SFT) loss with model parameters \(\boldsymbol{\theta}\).  
The empirical training loss over the dataset \(\mathcal{S}\) is then defined as:
\[
L_{\mathcal{S}}(\boldsymbol{\theta}) \triangleq \frac{1}{N} \sum_{j=1}^N \ell(\boldsymbol{\theta}; s_j, i_j).
\]
\vspace{-15pt}
\subsection{IF Definition in LRS}
\label{sec:if_defition}
While the LRS focuses on predicting the next item given a sequence, it is also important to ensure that items from different groups are treated in a balanced and representative manner.  
Otherwise, the LRS may reinforce historical biases by disproportionately favoring items from dominant groups.  
In sequential recommendation, fairness aims to maintain a balanced recommendation pattern across item groups with respect to their historical presence.  
Rather than focusing on the predictive utility of each group, we evaluate fairness by comparing how the recommendation distribution deviates from the historical distribution across groups, following prior works such as~\cite{steck2018calibrated,lincrank,jiang2024item}. 

Let the item groups be denoted as \(G_\mathcal{I} = \{G_1, G_2, \dots, G_N\}\).  
For each group \(G_n\), we define its historical ratio as the proportion of items from \(G_n\) that appear in users’ historically interacted items.  
This ratio reflects the natural prevalence of items in \(G_n\) within the observed data.  
During recommendation, given a set of input sequences \(\mathcal{S}\), the model generates top-$K$ recommendation lists.  
We define the recommended ratio of group \(G_n\) as the proportion of items from \(G_n\) appearing among all recommended items within these lists.  

To achieve fair recommendation, the model should maintain consistent relative recommendation intensity across groups.  
In other words, the ratio between the recommended proportion and the historical proportion should be similar for all groups.  
If this ratio varies considerably between groups, the model would disproportionately recommend certain groups relative to their historical representation, resulting in unfair recommendation.

Formally, we define $\epsilon$-Item-side Fairness in LRS as follows:

\textbf{Definition 1 ($\epsilon$-Item-side Fairness).}  
A LRS satisfies $\epsilon$-item-side fairness if, for any two item groups \(G_a, G_b \in G_\mathcal{I}\),  
the difference between their relative recommendation ratios is bounded by a threshold $\epsilon$:
\[
\Big| 
\frac{\mathrm{Ratio}^{\mathrm{rec}}(G_a)@K}{\mathrm{Ratio}^{\mathrm{hist}}(G_a)} -
\frac{\mathrm{Ratio}^{\mathrm{rec}}(G_b)@K}{\mathrm{Ratio}^{\mathrm{hist}}(G_b)}
\Big| \leq \epsilon,
\]
where $\mathrm{Ratio}^{\mathrm{rec}}(G_n)@K$ denotes the empirical proportion of items from \(G_n\) appearing in top-$K$ recommendations,  
and $\mathrm{Ratio}^{\mathrm{hist}}(G_n)$ represents its historical proportion.  
A smaller $\epsilon$ indicates stronger fairness, implying that the recommendation distribution across item groups remains well aligned with their historical distributions.

The motivation for adopting a fairness definition based on aligning the proportion of recommended items with users’ historical interaction distribution, rather than enforcing equal utility or equal exposure across groups, is twofold. First, defining utility solely according to the target item’s group is inappropriate, since the input sequence often contains items from multiple groups and therefore cannot fully capture potential unfairness. Second, equal exposure fails to account for user preferences. Recommending items from different groups with equal probability may deteriorate recommendation quality. This could lead to certain items being recommended to users with little or no interest in them, ultimately degrading user experience.
\section{Causes of Unfairness in LRSs}
\label{sec:cause}

\subsection{Compositional Unfairness in LRSs}
For an item $i \in G_n$,  
the prediction score of the LRS after SFT can be written as
\begin{equation}
f_{ft}(s,i) = \tilde f(s,i) + b^{\mathrm{total}}(G_n),
\end{equation}
where $\tilde f(s,i)$ denotes the ideal unbiased preference score.  
The group-level bias term $b^{\mathrm{total}}(G_n)$ represents the 
systematic deviation of the model’s scoring pattern for items from group $G_n$.
This bias arises from two sources:  
(1) the \textit{inherited bias from pre-training}, originating from the general-domain corpus used to train the backbone LLMS; and  
(2) the \textit{data-driven bias from SFT}, introduced by the downstream recommendation data and optimization objective.  
SFT therefore produces a \textit{composite} unfairness by jointly 
retaining pretraining preferences and adapting to new distributions.
Since our analysis focuses on group-level fairness, we do not explicitly model intra-group bias or item-specific deviations within each group.
All disparities are considered at the aggregated group level, which captures the dominant sources of unfairness in LRSs.

Assuming a softmax-like exposure RS and that the unbiased baseline follows 
the historical ratios $\mathrm{Ratio}^{\mathrm{hist}}(G_n)$ defined in Section~\ref{sec:if_defition},  
the expected recommendation ratio of group $G_n$ is
\begin{equation}
\mathrm{Ratio}^{\mathrm{rec}}(G_n)
=
\frac{\mathrm{Ratio}^{\mathrm{hist}}(G_n)\,
\exp(b^{\mathrm{total}}(G_n))}
{\sum_{m}\mathrm{Ratio}^{\mathrm{hist}}(G_m)\,
\exp(b^{\mathrm{total}}(G_m))}.
\end{equation}
When all group-level biases are equal, the recommendation distribution perfectly matches the historical one.

\subsection{Interaction Between Pre-training and SFT}

The relative recommendation intensity of each group is defined as
$
R(G_n) \triangleq
\frac{\mathrm{Ratio}^{\mathrm{rec}}(G_n)}
{\mathrm{Ratio}^{\mathrm{hist}}(G_n)}
$.
For any two groups $G_a$ and $G_b$,
\begin{equation}
\label{equ:R_group}
\frac{R(G_a)}{R(G_b)} 
= 
\exp\!\big(b^{\mathrm{total}}(G_a) - b^{\mathrm{total}}(G_b)\big).
\end{equation}
Eq.\eqref{equ:R_group} shows that group-level deviation is completely determined 
by the difference in total biases.  
In downstream fine-tuning, $b^{\mathrm{total}}$ reflects 
both the pre-training-induced prior and the adaptation to downstream recommendation data, 
which are often entangled and interact in nontrivial ways.

To describe this interaction, we decompose
\[
b^{\mathrm{total}}(G_n)
\triangleq b^{p}(G_n) + \delta(G_n),
\]
where $b^{p}(G_n)$ denotes the pretraining-induced bias 
and $\delta(G_n)$ is the bias shift induced during SFT.
The shift $\delta(G_n)$ is influenced simultaneously by 
the recommendation data distribution and the inherited pretraining preferences.  
It can therefore either amplify or attenuate existing disparities.

We define the centered biases
$\Delta^{p}(G_n) \triangleq b^{p}(G_n) - \bar b^{p}$,$
\Delta^{\delta}(G_n) \triangleq \delta(G_n) - \bar \delta$,
where averages are taken over $\mathrm{Ratio}^{\mathrm{hist}}(G_n)$.
Under a small-bias approximation,
\[
\log R(G_n) \approx \Delta^{p}(G_n) + \Delta^{\delta}(G_n),
\]
and the dispersion of group intensities satisfies
\begin{equation}
\mathrm{Var}[\log R(G_n)]
\approx
\mathrm{Var}(\Delta^{p})
+\mathrm{Var}(\Delta^{\delta})
+2\,\mathrm{Cov}(\Delta^{p},\Delta^{\delta}).
\end{equation}
The first term quantifies the inherent disparity inherited from pretraining, 
while the second measuring the additional group-level dispersion introduced during SFT, conditioned on the pre-training initialization. The covariance term determines whether SFT mitigates 
($\mathrm{Cov}<0$) or amplifies ($\mathrm{Cov}>0$) the inherited unfairness.

\subsection{Estimating Group Unfairness}
To quantify fairness empirically, we first estimate biases at the item level and then aggregate them to the group level.  
Inspired by the line of work on logit adjustment methods~\cite{menonlong,zhu2023generalized},  
we introduce a calibration-based formulation to characterize bias in the logit space.  
Let $q=(q_1,\ldots,q_{|\mathcal{I}|})$ be a probability simplex over all items  
($q_i\ge0$, $\sum_i q_i=1$), representing the calibration weights assigned to individual items.

\textbf{Item-level calibration.}  
Given the zero-shot base LRS without SFT $f_{zs}$ and a validation distribution $P_t$,  
we estimate the calibration vector by solving
\begin{equation}
q^{*}
=
\arg\min_{q}
\mathcal{R}_t\!\big(f_{zs}-\log q\big)
\quad
\mathrm{s.t. } q_i\ge0,\ \sum_i q_i=1,
\end{equation}
where $\mathcal{R}_t(\cdot)$ denotes the expected validation loss.  
The resulting $\hat b^{p}(i)=\log q_i^{*}$ gives the pretraining bias for each item.

\textbf{Fine-tuning bias shift.}  
Applying the same calibration to the fine-tuned model $f_{ft}$ yields $q_{ft}^{*}$,  
and the item-level bias shift is computed as $\hat{\delta}(i)
=
\log q_{ft,i}^{*}
-
\hat{b}^{p}(i).$

\textbf{Group-level aggregation.}  
Let $\omega_i$ denote the relative weight of item $i$ within its group $G_n$,  
satisfying $\sum_{i \in G_n} \omega_i = 1$.  
Weights $\omega_i$ can be defined based on the historical exposure distribution or validation sample frequency,  
ensuring consistency with the group-level centering used above.  
The group-level biases are then computed as
\begin{equation}
\hat b^{p}(G_n)
=
\sum_{i \in G_n} \omega_i \, \hat b^{p}(i),
\qquad
\hat\delta(G_n)
=
\sum_{i \in G_n} \omega_i \, \hat\delta(i).
\end{equation}
This weighted aggregation preserves fine-grained item-level calibration while maintaining alignment with exposure-weighted group statistics. 

\textbf{Empirical study.}
\begin{figure}
    \centering
    \subfigure{
        \includegraphics[width=0.46\linewidth, trim=12 15 18 14, clip]{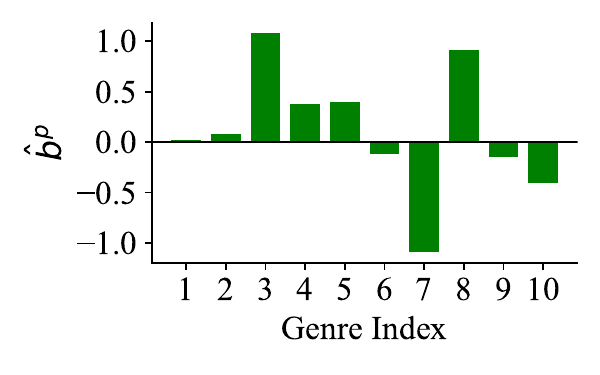}
    }
    \subfigure{
        \includegraphics[width=0.46\linewidth, trim=12 15 18 14, clip]{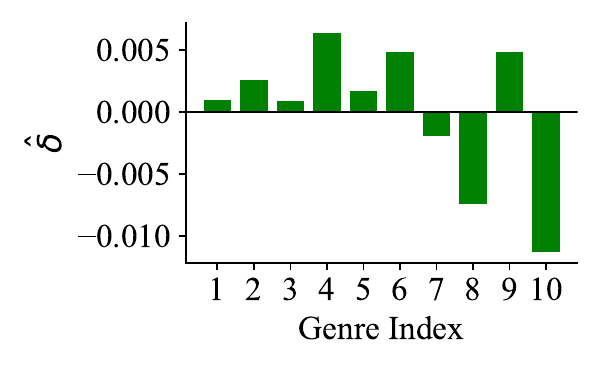}
    }
    \caption{Group-level Bias Estimation on ML-1M.}
    \label{fig:bias_estm}
    \vspace{-10pt}
\end{figure}
We estimate Genre fairness of LLama2-7b on the ML-1M dataset. Following the experimental procedure of \cite{gao2025sprec}, we select 10 genre groups and sample 4,096 instances. We then perform fairness estimation on both the original LLM and the SFT-tuned LLM, respectively.
Results are shown in Fig.~\ref{fig:bias_estm}. The index–genre mapping is provided in the Appendix~A.
We observe that the pre-training centered biases $\Delta^{p}(G_n)$ exhibit large dispersion. Besides the inherent genre-related preferences embedded in the LLM, this is mainly because the raw outputs of the base LLM without SFT deviate substantially from the desired item-level predictions, often producing long and irrelevant text rather than focused recommendation signals. Such behavior leads to extremely high variance in the induced scores and consequently produces strong apparent bias at the group level.

Among the 10 groups, only indices 6, 8, and 9 show bias directions inconsistent with pre-training, while the remaining groups follow the same trend. Together with the final covariance estimate $\mathrm{Cov}=7.73\times10^{-4} > 0$, this indicates that SFT generally aligns with the inherited pre-training tendencies. As a result, the SFT stage reinforces rather than corrects these disparities, ultimately amplifying group-level unfairness.

In Summary, group-level unfairness in LRSs originates from a composite bias 
that is shaped by both the pre-training and SFT stages. Pre-training establishes the initial bias pattern, while SFT transforms it under the influence of downstream data. Together, these two factors accumulate and ultimately amplify the resulting unfairness.

\section{Unfair-to-Fair Evolving}
\begin{figure*}[t]
    \centering
    \includegraphics[width=0.9\linewidth, trim=4.1cm 4.2cm 5.5cm 3.6cm, clip]{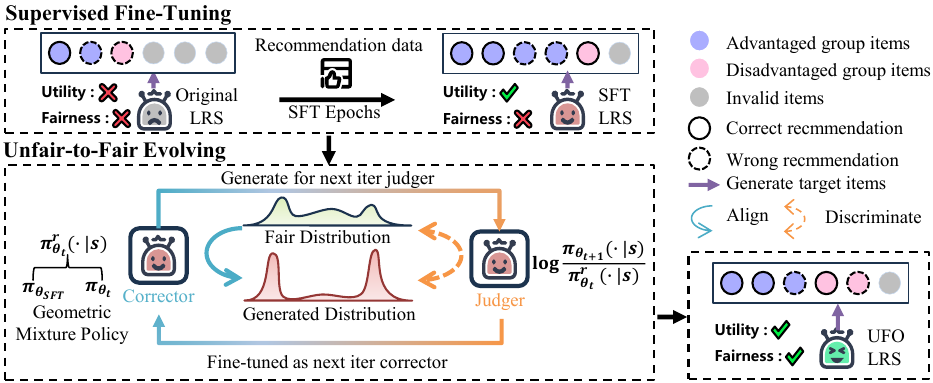}
    % \vspace{-10pt}
    \caption{An illustration of the pipelines of SFT and our proposed \model{}. SFT tends to amplify the inherent unfairness in LRSs introduced during pre-training. In contrast, \model{} aims to mitigate unfairness within LRSs completely. By iteratively identifying and enhancing fairness through the self-play of a \textit{judger} and a \textit{corrector}, \model{} enables LRS to evolve from an unfair state to a fair one.}
    \label{fig:frame}
    \vspace{-10pt}
\end{figure*}

Noting the complex interplay between pre-training and SFT in shaping unfairness in LRSs, methods that target fairness improvement at only one stage are inherently limited.
The analysis in Section~\ref{sec:cause} inspires us to address the unfairness issue holistically by directly calibrating the final outputs of the post-SFT LRS in a single step.

To this end, we introduce \model{}, a unified method designed to directly address all forms of unfairness in LRSs, regardless of whether this unfairness originates from pre-training or SFT.
Rather than meticulously adjusting the SFT process, \model{} directly aligns the post-SFT LRS, with a fair objective.
Consequently, \model{} requires a fair distribution as the optimization target.

However, there is no prior knowledge for the model to generate an ideal fair distribution directly. Moreover, such a fair distribution must also maintain the model's performance. 
Thus, it is intuitive to consider utilizing the recommendation dataset used during SFT.
Although the recommendation datasets (i.e., consisting of constructed recommendation prompt and result pairs) exhibit popularity bias. Such bias does not constitute unfairness. It merely influences the recommendation model, resulting in certain groups not achieving the ideal recommendation performance expected in the dataset~\cite{abdollahpouri2020connection}.
These datasets still represent a distribution that ideally balances performance and fairness.

Consequently, we adopt the recommendation dataset utilized for SFT as a proxy for the fair distribution in our proposed framework. 
Let $\mathcal{S} = \{(s_j, i_j)\}_{j=1}^N$ denote the recommendation dataset,  
where $s_j$ represents the input sequence of previously interacted items,  
and $i_j$ denotes the corresponding target item to be recommended.  
In this setting, $s$ is sampled from the sequence distribution $\chi$,  
and $i$ is drawn from the conditional distribution $p_{\mathcal{S}}(i|s)$,  
which serves as a proxy for the fair distribution $p_{\mathrm{fair}}(i|s)$ in our context.
\vspace{-5pt}

\subsection{Overview}
Inspired by the strong performance of the self-play mechanism across various domains~\cite{goodfellow2014generative,silver2017mastering,munos2023nash} and its successful applications in LLMs~\cite{chen2024self}, we leverage the self-play mechanism to address unfairness in LRSs. As illustrated in Fig.~\ref{fig:frame}, we consider the enhancement of fairness in LRS as a two-player game. 
In this game, we introduce two core players: a \textit{judger} and a \textit{corrector}. The judger's objective is to identify unfairness inherent within LRS's responses and to accurately discriminate between unfair and fair responses. 
The corrector's primary goal is to improve its own fairness, striving to generate responses that are as similar as possible to fair responses while preserving recommendation performance. 
The corrector represents the LRS from previous iterration, whereas the judger is the LRS being trained in the current iteration. 
Both players originate from different iterations of the same LRS, constructing the self-play mechanism. 

Moreover, we introduce two key components in our framework: \textit{Distributional next-item generation} and a \textit{geometric mixture policy}.  
The former explicitly models the group-wise distribution of items generated by the LRS, which facilitates the identification and mitigation of unfairness.  
The latter serves to preserve recommendation performance by forming a mixture between the LRS from the previous iteration and the original LRS.  
Since the original LRS exhibits superior recommendation accuracy, incorporating this mixture term into the optimization constraint helps maintain the utility.
% This geometric mixture is also employed in ~\cite{munos2023nash}. \zjm{need this?}
%
We consider every iteration of \model{} as a self-play iteration, with a particular self-play iteration denoted as $t$. 
In a new self-play iteration $t+1$, we need to sequentially update the judger and the corrector.
\vspace{-10pt}
\subsection{Distributional next-item generation and adaption}
LRSs formulate sequential recommendation as a next-item generation task.  
To better expose and calibrate potential group-level biases without altering the model’s inherent knowledge,  
we introduce a \textit{Distribution-aware next-item generation} paradigm that reformulates the original next-item prediction process by explicitly tagging each generated item with its group identity.  
This approach retains the original model outputs while making the group-wise distribution of generations observable and calibratable.  

Leveraging the language modeling nature of LRSs, we factorize the next-item generation process into two components:  
(1) the \textit{group-level distribution} $\pi_{\boldsymbol{\theta}}(g|s)$, which estimates the probability that the next item belongs to each group given the input sequence $s$; and  
(2) the \textit{item-level generation} $\pi_{\boldsymbol{\theta}}(i|g,s)$, which predicts the specific next item $i$ conditioned on both the group $g$ and the input sequence.  
Formally,
\begin{equation}
\pi_{\boldsymbol{\theta}}(i|s)
=
\sum_{g \in \mathcal{G}} 
\pi_{\boldsymbol{\boldsymbol{\theta}}}(i|g,s)\, \pi_{\boldsymbol{\theta}}(g|s),
\label{eq:distributional_next_item}
\end{equation}
where $\mathcal{G}$ denotes the set of groups.  
This formulation decouples the model’s implicit preference over group membership from item-level generation,  
allowing clearer analysis and correction of group-level biases.  

In implementation, we instantiate group tokens $[\text{G0}], [\text{G1}], \ldots$ to represent specific groups (e.g., $g_0 = [\text{G0}]$).  
During training, each generated item from the original LRS is annotated with its corresponding group token,  
and the model is trained to reproduce the same outputs under this new distributional format.  
We align the predicted group-token probabilities with the group distribution implied by the original model outputs using a Kullback–Leibler (KL) divergence loss:$ \mathcal{L}_{\mathrm{KL}}
=
\mathrm{KL}\big(\pi_{\boldsymbol{\theta}}(g|s)\,\|\,\pi_{\mathrm{orig}}(g|s)\big)$.
To ensure consistent item prediction, we further include a negative log-likelihood (NLL) loss for predicting the correct group token, and another NLL loss for maintaining the original item-generation behavior: $\mathcal{L}_{\mathrm{group}}
= - \mathbb{E}_{(s,g)} \log \pi_{\boldsymbol{\theta}}(g|s)$,
$\mathcal{L}_{\mathrm{item}}
= - \mathbb{E}_{(s,i,g)} \log \pi_{\boldsymbol{\theta}}(i|g,s)$. The final objective is given by
$
\mathcal{L}_{\mathrm{total}}
=
\mathcal{L}_{\mathrm{item}}
+
\lambda_1 \mathcal{L}_{\mathrm{group}}
+
\lambda_2 \mathcal{L}_{\mathrm{KL}}
$, where $\lambda_1$ and $\lambda_2$ balance generation accuracy and format consistency.  
This post-training adaptation modifies only the output representation while preserving the model’s learned knowledge and original recommendation ability,  
thus enabling interpretable group-level calibration with minimal computational overhead.

In addition, this formulation naturally enhances generation diversity. For instance, users with similar historical sequences may still exhibit preferences for different groups of next items.  
Compared with the conventional next-item generation paradigm, the distributional formulation requires only a single generation step during inference, meaning that the model directly outputs the most probable next item rather than the full group distribution, while introducing only a minimal number of additional tokens.
\subsection{Judger Step: Identify Unfairness}
Although we have redefined $\pi_{\boldsymbol{\theta}}(i|s)$ under the distributional next-item generation framework,  
this reformulation does not affect the outputs of the LRS or the definition of $p_{\mathrm{fair}}(i|s)$.  
Each item is associated with exactly one group — for the LRS, this corresponds to the group with the highest predicted probability,  
while for $p_{\mathrm{fair}}$ it can be directly determined from data annotations.  
Therefore, for simplicity, we continue to use $\pi_{\boldsymbol{\theta}}(i|s)$ and $p_{\mathrm{fair}}(i|s)$ to denote the distributional next-item generation process,  
with the understanding that $s$ implicitly includes the corresponding group label.

The judger's objective is to identify unfairness, which is achieved by distinguishing between the fair distribution $p_{\mathrm{fair}}$ and the unfair distribution generated by the corrector from the last iteration $t$. To achieve the judger's objective, we denote the corrector from the last iteration as $\pi_{\boldsymbol{\theta}_t}^{'}$.
Through fine-tuning in the current iteration $t$, we aim to obtain $\pi_{\boldsymbol{\theta}_{t+1}}$ such that $f_{\pi_{\boldsymbol{\theta}_{t+1}}}$ serving as the judger can maximize the expected value gap between $p_{\mathrm{\mathrm{fair}}}$ and the distribution of responses generated by the corrector from the last iteration  $\pi_{\boldsymbol{\theta}_t}^{'}$:
\begin{align}
\label{equ:maxf}
\pi_{\boldsymbol{\theta}_{t+1}} = \arg\max_{\pi_{\boldsymbol{\theta}}} 
\mathbb{E} \left[ f_{\pi_{\boldsymbol{\theta}}}(s, i) - f_{\pi_{\boldsymbol{\theta}}}(s, i') \right],
\end{align}
where $s \sim \chi$, $i \sim p_{\mathrm{fair}}(\cdot|s)$ and $i' \sim \pi_{\boldsymbol{\theta}_t}^{'}(\cdot|s)$. 

Specifically, given \(\pi_{\boldsymbol{\theta}_{t+1}}\) and the function \(f_{\pi_{\boldsymbol{\theta}_{t+1}}}\) along with a sequence of target item responses \(i\) to the input \(s\), the value of \(f_{\pi_{\boldsymbol{\theta}_{t+1}}}(s, i)\) reflects how the judger believes that \(i\) originates from the fair distribution rather than an unfair one. 
This quantifies the degree of unfairness present in \(i\) that requires correction and measures the distance between the current response and the fair objective.

To formalize the objective in a more general optimization framework, we can rewrite Eq.\eqref{equ:maxf} within a loss function $\ell(\cdot)$, which is both monotonically decreasing and convex:
\begin{align}\label{equ:mainloss}
\pi_{\boldsymbol{\theta}_{t+1}} = \arg\min_{\pi_{\boldsymbol{\theta}}} 
\mathbb{E}[ \ell(f_{\pi_{\boldsymbol{\theta}}}(s, i) - f_{\pi_{\boldsymbol{\theta}}}(s, i') )].
\end{align}
\subsection{Corrector Step: Correct Unfairness while Maintaining Recommendation Performance}
The corrector's objective is to generate responses that are more fair, that is, more similar to those generated by \( p_{\mathrm{fair}} \) without impairing recommendation performance. Ideally, the updated parameters of the judger \( \pi_{\boldsymbol{\theta}_{t+1}} \), should result in high values for \( f_{\pi_{\boldsymbol{\theta}_{t+1}}} \) when given \( i \sim p_{\mathrm{fair}}(\cdot|s) \), and low values when given \( i' \sim \pi_{\boldsymbol{\theta}_t}^{'}(\cdot|s) \).
% where \( \pi_{\boldsymbol{\theta}_t}^{\mathrm{r}} \) denotes the regularized policy of the corrector from the previous iteration $t$. 
%
The update of the corrector is based on the judger. However, overly focusing on improving performance on the judger can neglect the assurance of recommendation performance. Therefore, we introduce the geometric mixture policy $\pi_{\boldsymbol{\theta}_t}^{\mathrm{r}}$ representing a geometric mixture between the older corrector $\pi_{\boldsymbol{\theta}_t}$ and the original corrector $\pi_{\boldsymbol{\theta}_\mathrm{SFT}}$:
\begin{align}
    \pi_{\boldsymbol{\theta}_t}^{\mathrm{r}}(i|s)=\frac{\pi_{\boldsymbol{\theta}_t}(i|s)^{1-\alpha} \ \pi_{\boldsymbol{\theta}_\mathrm{SFT}}(i|s)^{\alpha}}{\sum_{i'}\pi_{\boldsymbol{\theta}_t}(i'|s)^{1-\alpha} \ \pi_{\boldsymbol{\theta}_\mathrm{SFT}}(y'|s)^{\alpha}} ,
\end{align}
% \begin{align}
%     \pi_{\boldsymbol{\theta}_t}^{\mathrm{r}}(y|x) \propto (\pi_{\boldsymbol{\theta}_t}(y|x))^{1-\alpha} \ (\pi_{\boldsymbol{\theta}_\mathrm{SFT}}(y|x))^{\alpha},
% \end{align}
where $\alpha \in [0,1]$ is a mixture parameter to balance $\pi_{\boldsymbol{\theta}_t}$ and $\pi_{\boldsymbol{\theta}_\mathrm{SFT}}$. This geometric mixture ensures that the LRS in our fine-tuning process remains close to $\pi_{\boldsymbol{\theta}_\mathrm{SFT}}$ with strong recommendation performance. We can rewrite Eq.\eqref{equ:mainloss} as:
\begin{align}
\pi_{\boldsymbol{\theta}_{t+1}} = \arg\min_{\pi_{\boldsymbol{\theta}}} 
\mathbb{E}[ \ell(f_{\pi_{\boldsymbol{\theta}}}(s, i) - f_{\pi_{\boldsymbol{\theta}}}(s, i') )],
\end{align}
where $s \sim \chi$, $i \sim p_{\mathrm{fair}}(\cdot|s)$ and $i' \sim \pi_{\boldsymbol{\theta}_t}^{\mathrm{r}}(\cdot|s)$.

Consequently, the corrector for the new iteration seeks to optimize towards a target that generates responses more closely resembling those of \( p_{\mathrm{fair}} \) while maintaining recommendation performance. Drawing on RL fine-tuning methods~\cite{bai2022training,rafailov2024direct}, we can formalize our objective as:
\begin{equation}
\label{equ:corrector}
\begin{aligned}
    \pi'_{\boldsymbol{\theta}_{t+1}} = \arg\max_{\pi_{\boldsymbol{\theta}}} \mathbb{E}_{s \sim \chi, i \sim \pi_{\boldsymbol{\theta}}(\cdot|s)} [f_{\pi_{\boldsymbol{\theta}_{t+1}}}(s,i)] \\
    - \beta \mathbb{E}_{s \sim \chi} [\mathrm{KL}(\pi_{\boldsymbol{\theta}}(\cdot|s) \parallel \pi_{\boldsymbol{\theta}_t}^{\mathrm{r}}(\cdot|s))],
\end{aligned}
\end{equation}
where \( \beta > 0\) is the regularization parameter. The KL-constrained objective ensures that the corrector does not diverge too far from the reference model during optimization, thereby guaranteeing the stability of fine-tuning. 
Recall that the introduced geometric mixture policy computes as $\pi_{\boldsymbol{\theta}_t}^{\mathrm{r}}(i|s) \propto (\pi_{\boldsymbol{\theta}_t}(i|s))^{1-\alpha} \ (\pi_{\boldsymbol{\theta}_\mathrm{SFT}}(i|s))^{\alpha}$. 
Thus, the original KL term in Eq.\eqref{equ:corrector} is rewritten as:
\begin{equation}
\label{equ:KL}
\begin{aligned}
   \mathrm{KL}(\pi_{\boldsymbol{\theta}}(\cdot|s) \parallel \pi_{\boldsymbol{\theta}_t}^{\mathrm{r}}(\cdot|s)) =  (1-\alpha)\mathrm{KL}(\pi_{\boldsymbol{\theta}}(\cdot|s) \parallel \pi_{\boldsymbol{\theta}_t}(\cdot|s)) \\
   + \alpha\mathrm{KL}(\pi_{\boldsymbol{\theta}}(\cdot|s) \parallel \pi_{\boldsymbol{\theta}_{\mathrm{SFT}}}(\cdot|s)) + c(s),
\end{aligned}
\end{equation}
where $c(s)$ is a normalization term independent of $\pi_{\boldsymbol{\theta}}$. Compared to a conventional self-play framework, the newly added KL regularization term for $\pi_{\boldsymbol{\theta}_{\mathrm{SFT}}}$ ensures that it remains as close as possible to the original LRS.

The KL-constrained optimization process in Eq.\eqref{equ:corrector} can be further expressed in the following form: 
\begin{align}
    \pi'_{\boldsymbol{\theta}_{t+1}}(i|s) \propto \pi_{\boldsymbol{\theta}_t}^{\mathrm{r}}(i|s) \exp (\beta^{-1}f_{\pi_{\boldsymbol{\theta}_{t+1}}}(s,i)).
\end{align}
In this form, the optimal update can be directly observed by steering the current corrector in the direction that improves fairness guided by the judger $f_{\pi_{\boldsymbol{\theta}_{t+1}}}$ towards an updated corrector that maximizes relative to the geometric mixture policy $\pi_{\boldsymbol{\theta}_t}^{\mathrm{r}}$.
%
% We can derive the optimal policy $\pi^{*}$ for the KL-constrained objective as:
% \begin{align}
%     \pi^{*}(y|x) = \frac{\pi_{\boldsymbol{\theta}_t}(\cdot|x) \exp (\beta^{-1}f_{\pi^{t+1}}(x,y))}{Z_{\pi_{\boldsymbol{\theta}_t}}(x)},
% \end{align}
% where the partition function is defined as $Z_{\pi_{\boldsymbol{\theta}_t}}(x) = \sum_{y} \pi_{\boldsymbol{\theta}_t}(y|x) \exp ( \beta^{-1} f_{\pi^{t+1}}(x,y) )$. 
Then, we can rewrite $f_{\pi_{\boldsymbol{\theta}_{t+1}}}$ using the optimal corrector $\pi'_{\boldsymbol{\theta}_{t+1}}$:
\begin{align}\label{equ:ft+1}
f_{\pi_{\boldsymbol{\theta}_{t+1}}}(s, i) = \beta \log \frac{\pi'_{\boldsymbol{\theta}_{t+1}}(i|s)}{\pi_{\boldsymbol{\theta}_t}^{\mathrm{r}}(i|s)} + \beta \log Z(s), 
\end{align}
where the partition function $Z(s)$ is defined as $Z(s) = \sum_{i} \pi_{\boldsymbol{\theta}_t}^{\mathrm{r}}(i|s) \exp ( \beta^{-1} f_{\pi_{\boldsymbol{\theta}_{t+1}}}(s,i) )$. 
%
% The partition function is a function of only $x$ and the policy $\pi_{\boldsymbol{\theta}_t}^{\mathrm{r}}$, but does not depend on $\pi_{\boldsymbol{\theta}_{t+1}}$.
%
We observe that the training objective of the judger relies solely on the discrepancy between two \( f_{\pi_{\boldsymbol{\theta}}} \) values. By substituting Eq.\eqref{equ:ft+1} into Eq.\eqref{equ:mainloss}, the partition function cancels, allowing us to express the fairness metric function exclusively in terms of the optimal corrector \( \pi'_{\boldsymbol{\theta}_{t+1}} \) and the geometric mixture policy of the corrector from last iteration \( \pi_{\boldsymbol{\theta}_t}^{\mathrm{r}} \). Consequently, we obtain \( \pi'_{\boldsymbol{\theta}_{t+1}} = \pi_{\boldsymbol{\theta}_{t+1}} \), which indicates that \( \pi_{\boldsymbol{\theta}_{t+1}} \) learned from Eq.\eqref{equ:mainloss} exactly represents the optimal choice for the corrector at iteration $t+1$.

\begin{algorithm}[tb]
\caption{UFO: Unfair-to-Fair Evolving}
\label{alg:ufo}
\textbf{Input}: Recommendation Dataset $\mathcal{S} =\{(s, i)\}_{i=1}^{n}$, Iterations $T$, Original LRS $\pi_{\boldsymbol{\theta}_\mathrm{SFT}}$.\\
\textbf{Output}: Final LRS $\pi_{\boldsymbol{\theta}_T}$.

\begin{algorithmic}[1] %[1] enables line numbers
\STATE Initialize $\pi_{\boldsymbol{\theta}_1} = \pi_{\boldsymbol{\theta}_\mathrm{SFT}}$.
\FOR{$t$ in $1 \dots T$}
    \STATE Get $\pi_{\boldsymbol{\theta}_t}^{\mathrm{r}}(i|s) \propto (\pi_{\boldsymbol{\theta}_t}(i|s))^{1-\alpha} \ (\pi_{\boldsymbol{\theta}_\mathrm{SFT}}(i|s))^{\alpha}$.
    \COMMENT{\textit{Corrector Step}}
    \STATE Generate $\{s_j,i_j,i'_j\}$ with $(s_j,i_j) \sim \mathcal{D}$ and $i'_j \sim \pi_{\boldsymbol{\theta}_t}^{\mathrm{r}}(\cdot|s_j)$.
    \COMMENT{\textit{Judger Step}}
    \STATE $\pi_{\boldsymbol{\theta}_{t+1}} \gets \pi_{\boldsymbol{\theta}_{t}} - \eta \nabla_{\boldsymbol{\theta}} L_\mathrm{UFO}(\pi_{\boldsymbol{\theta}};\pi_{\boldsymbol{\theta}_t}^{\mathrm{r}})$.
\ENDFOR
\STATE \textbf{return} Final LRS $\pi_{\boldsymbol{\theta}_T}$
\end{algorithmic}
\end{algorithm}

% \subsubsection{Iteratively Unfair-to-Fair Optimization Pipeline}
\subsection{Putting Together}
Integrating the above two steps for a parametrized policy $\pi_{\theta}$, our training objective becomes:
\begin{align}
\label{loss:UFO}
    \mathcal{L}_{\mathrm{UFO}}(\pi_{\theta};\pi_{\boldsymbol{\theta}_t}^{\mathrm{r}}) = \mathbb{E} [\ell(\beta \log \frac{\pi_{\theta}(i|s)}{\pi_{\boldsymbol{\theta}_t}^{\mathrm{r}}(i|s)} - \beta \log \frac{\pi_{\theta}(i'|s)}{\pi_{\boldsymbol{\theta}_t}^{\mathrm{r}}(i'|s)})],
\end{align}
where $s \sim \chi$, $i \sim p_{\mathrm{fair}}(\cdot|s)$ and $i' \sim \pi_{\boldsymbol{\theta}_t}^{\mathrm{r}}(\cdot|s)$. Thus, we can iteratively refine the LRS through self-play based on the LRS from the previous iteration. The entire process of \model{} is summarized in Algorithm \ref{alg:ufo}. 

In iteration $t$, we first obtain the recommendation results generated by the geometric mixture of the corrector from the previous iteration. We provide more details of generating recommendation results after mixing in Appendix~B.
These results, together with the recommendation dataset, are then used for the next iteration $t+1$. In iteration $t+1$, the judger is trained using the LRS's parameters of the corrector from the last iteration, yielding updated LRS's parameters for the judger in the current iteration while not deviating too much from original LRS due to the geometric mixture.
The corrector then directly employs these freshly trained LRS's parameters to generate more fair recommendation results. These updated parameters are subsequently fed back into the fine-tuning of the judger for the next iteration. This iterative optimization enables the LRS to progressively transition from an unfair state towards a fair one while maintaining recommendation performance. Notably, in Appendix~C, 
 we show that \model{} achieves global optimum if and only if geometric mixture policy $\pi_{\boldsymbol{\theta}_t}^{\mathrm{r}}$ generates ideally fair recommendation results, which aligns with the target fair distribution $p_{\mathrm{fair}}$.

% SPRec~\cite{gao2025sprec} also proposes to use self-generated samples, but its approach is fundamentally different from that of \model{}.  
% Specifically, SPRec aims to mitigate the bias inherent in DPO by introducing an iterative DPO framework that leverages self-generated responses as negative samples.  
% In contrast, \model{}, similar in spirit to the Generative Adversarial Network (GAN)~\cite{goodfellow2014generative}, Wasserstein GAN~\cite{arjovsky2017wasserstein} and related IPM-based approaches~\cite{zhou2023natural,roch2025reduction,chen2024self},  
% maximizes the Integral Probability Metric~\cite{muller1997integral}, whereas SPRec is grounded in the Bradley–Terry (BT) model.  
% Moreover, \model{} reformulates the generation process into a distributional item generation paradigm.

\section{Experiments}
\subsection{Experimental Settings}
\subsubsection{Datastes}
We conduct experiments on three real-world datasets that are widely used in recommendation tasks. 
\textbf{ML-1M}\footnote{https://grouplens.org/datasets/movielens/}: The MovieLens dataset is one of the most extensively utilized datasets in recommender system research. ML-1M contains 1 million ratings. \textbf{Steam}\footnote{https://cseweb.ucsd.edu/~jmcauley/datasets.html}: The Steam dataset contains reviews from the Steam video game platform.
\textbf{ADM}\footnote{http://jmcauley.ucsd.edu/data/amazon/}: The Amazon Digital Music (ADM) dataset, includes ratings of digital music. The statistical details of datasets are summarized in Table~\ref{tab:dataset}.

To simulate real-world sequential recommendation scenarios, we adopt the pre-processing procedure from sequential recommendation tasks, constructing sequences of length 10 based on the timestamps of item interactions. Following this, we partition these sequences into training, validation, and testing sets in an 8:1:1 ratio. To address the significant imbalance in the distribution of game genres within user interactions present in the Steam dataset, we refer to the approach in ~\cite{jiang2024item} and exclude games with fewer than 10,000 total interactions for their respective genres. For similar reasons, we also delete items from the ADM dataset that have fewer than 10 interactions.

To evaluate group fairness, we implement three item grouping strategies : based on popularity, genre, and title length. For popularity-based grouping, we rank items according to their respective interaction counts and divide them into five equally sized groups. In genre-based grouping, we assign items to groups directly based on the genres extracted from the dataset. For title length-based grouping, items are ranked by the length of their titles and then divide into five equally sized groups.

\begin{table}
\centering
\caption{Summary of datasets.}
\label{tab:dataset}
\begin{tabular}{lccc}  
\toprule
Dataset   & \# Items    & \# Interactions & \# Sequences \\
\midrule
ML-1M  & 3,883 & 1,000,209  & 939,809  \\
Steam  &  32,135 & 1,307,310   & 138,970 \\
ADM  &  266,414 & 836,006   & 22,982 \\
\bottomrule
\end{tabular}
\vspace{-10pt}
\end{table}

\subsubsection{LRS Model}
We select the representative BIGRec~\cite{bao2025bi} as the target LRS. BIGRec, a LRS built upon SFT, serves as the base model for several hybrid recommendation approaches, which can be viewed as extensions that refine and expand the BIGRec paradigm. 

Similar to original settings of BIGRec, we adopt Llama-2-7B as the base model. We first perform instruction fine-tuning and subsequently train the model on recommendation dataset.
In addition, we also incorporate other LRSs and base LLMs for comparison. Specifically, we include LLaRA~\cite{liao2024llara}, whose token space mixes the embeddings from traditional RS, where we adopt SASRec~\cite{kang2018self}. For the base LLMs, we further use Qwen2.5-7B-Instruct (Qwen2.5)~\cite{team2024qwen2}, which already possesses strong instruction-following capabilities and is directly fine-tuned on recommendation data.
% Similar to original settings of BIGRec, we adopt Llama-2-7B as the base model. We first perform instruction fine-tuning and subsequently train the model on 65536 sampled sequences without altering the original training dataset distribution (using all sequences from the ADM dataset because its size is less than 65536).
\subsubsection{Compared Methods}
% We compare our proposed \model{} with existing methods that are applicable for enhancing IF in LRSs.
\begin{itemize}[leftmargin=*]\setlength{\itemsep}{2pt}
    \item \textbf{RW}: Re-weighting~\cite{jiang2024item} improves IF by adjusting training weights based on the group popularity within the dataset, thereby enhancing IF along a specific dimension of the LRSs.
    \item \textbf{D3}: Debiasing-Diversifying Decoding~\cite{bao2024decoding}, guided by SASRec during the decoding process, aims to mitigate homogenization issues in recommendation outcomes and increase diversity. This approach can also be regarded as a strategy to improve IF.
    \item \textbf{RosePO}: RosePO~\cite{liao2024rosepo}, integrates negative sampling with personalized uncertainty modeling. It enhances LRS robustness and mitigates popularity bias, which contributes to improved popularity IF. Therefore, we include it in our comparison.
    \item \textbf{SPPO}: Self-play DPO~\cite{gao2025sprec} enhances DPO by leveraging self-generated data from the LRS as negative samples to mitigate the bias caused by imbalance in negative examples.
\end{itemize}

\subsubsection{Evaluation Metrics}
Following prior studies~\cite{jiang2024item,gao2025sprec} and our definition in Section~\ref{sec:if_defition}, we evaluate group-level unfairness by comparing the model’s recommendation distribution with the historical interaction distribution.  
Let $\mathrm{GP}(G)$ denote the proportion of recommended items belonging to group $G$, and $\mathrm{GH}(G)$ denote the corresponding historical proportion.  
The group unfairness for $G$ is defined as
$\mathrm{GU}(G) = \mathrm{GP}(G) - \mathrm{GH}(G)$,  
where $\mathrm{GU}(G) > 0$ indicates over-recommendation and $\mathrm{GU}(G) < 0$ indicates under-recommendation.  
To aggregate group disparities, we adopt two metrics:  
(1) \textbf{Mean Group Unfairness (MGU)} quantifies the overall level of unfairness,  
$\mathrm{MGU} = \frac{1}{|\mathcal{G}|}\sum_{G\in\mathcal{G}}|\mathrm{GU}(G)|$;  
and (2) \textbf{Disparity Group Unfairness (DGU)} measures the largest fairness gap among groups,  
$\mathrm{DGU} = \max_G \mathrm{GU}(G) - \min_G \mathrm{GU}(G)$.  
These metrics jointly capture both average and extreme group-level unfairness in LRSs.

Given that both grounding operation and beam search in BIGRec can introduce substantial additional factors affecting fairness~\cite{jiang2024item}, we focus on the fairness metric on the first item generated by BIGRec. This focus provides a clearer representation of the inherent fairness characteristics of LRSs.

Furthermore, to investigate the impact of \model{} on recommendation performance, we employ two widely used metrics, i.e., Normalized Discounted Cumulative Gain (NDCG@K) and Hit Ratio (HR@K), to evaluate the recommendation performance of the LRS on the testing set.

\subsection{Implement Details}
\subsubsection{Training Settings}
\paragraph{SFT Stage}
Following the training settings of BIGRec~\cite{bao2025bi} strictly, we conduct recommendation-specific SFT to adjust the base LLM for recommendation tasks. We utilize the Llama-2-7b model as the base model and perform Instruction Fine-tuning using the Alpaca-Cleaned dataset. Subsequently, we further fine-tune the model on the recommendation dataset, which is also formatted in the Alpaca style, to specialize it for recommendation tasks.

\paragraph{\model{} Generation Stage}
Following the training data volume settings of optimization methods such as DPO~\cite{rafailov2024direct}, we randomly sample \( m/2 \) examples from a dataset containing \( m \) samples at each iteration to provide to the geometric mixture policy for generation. Consequently, in each \model{} iteration, \( m/2 \) triplets are involved in the fine-tuning process.

\paragraph{\model{} Fine-tuning Stage}
Regarding the selection of \( \ell \) in Eq.\eqref{loss:UFO}, we adopt the logistic loss function \( \ell(t) := \log(1 + \exp(-t)) \). Such a selection is informed by its properties of non-negativity, smoothness, and exponential decay as \( t \to \infty \). The selected loss function aids in preventing the excessive growth of the absolute value of \( f \). Both our recommendation-specific SFT and \model{} utilize the LoRA approach.
\subsubsection{Hyper-parameters}
With the primary goal of maximizing fairness, we do not focus on enhancing recommendation performance but instead aim to ensure it remains no lower than the baseline. Specifically, we set \(\beta = 0.1\), utilize the AdamW optimizer with a learning rate of \(5.0 \times 10^{-7}\), and carry out one training epoch per iteration. In our main experiments, \(\alpha\) is set to 0.4, and the number of iterations ranges from 3 to 5, changing upon the specific dataset.

\begin{table*}[t]
\caption{Results in terms of IF (MGU and DGU) and recommendation performance (NDCG and HR) with different strategies for groups split. We highlight the top results in \textbf{bold}, and additionally report the relative improvements (\%) of \model{} over the best baseline.}
\label{tab:overview}
\centering
\resizebox{\textwidth}{!}{
\begin{tabular}{clcccccccc}
    \toprule
    \multirow{2}[4]{*}{Dataset} & \multicolumn{1}{c}{\multirow{2}[4]{*}{Method}} & \multicolumn{2}{c}{Popularity Fairness $\downarrow$} & \multicolumn{2}{c}{Genre Fairness $\downarrow$} & \multicolumn{2}{c}{Length Fairness $\downarrow$} & \multicolumn{2}{c}{Accuracy $\uparrow$} \\
\cmidrule{3-10}          &       & MGU & DGU & MGU & DGU & MGU & DGU & NDCG@5 & HR@5 \\
    \midrule
    \multirow{6}[2]{*}{ML-1M} & BIGRec & \multicolumn{1}{c}{0.0866 } & \multicolumn{1}{c}{0.3146 } & \multicolumn{1}{c}{0.0359 } & \multicolumn{1}{c}{0.2344 } & \multicolumn{1}{c}{0.0088 } & \multicolumn{1}{c}{0.0341 } & \multicolumn{1}{c}{0.0282 } & \multicolumn{1}{c}{0.0310 } \\
          & RW    & \multicolumn{1}{c}{0.0770 } & \multicolumn{1}{c}{0.2975 } & \multicolumn{1}{c}{0.0335 } & \multicolumn{1}{c}{0.2079 } & \multicolumn{1}{c}{0.0089 } & \multicolumn{1}{c}{0.0347 } & \multicolumn{1}{c}{0.0278 } & \multicolumn{1}{c}{0.0298 } \\
          & D3    & \multicolumn{1}{c}{0.0822 } & \multicolumn{1}{c}{0.3054 } & \multicolumn{1}{c}{0.0372 } & \multicolumn{1}{c}{0.2611 } & \multicolumn{1}{c}{0.0082 } & \multicolumn{1}{c}{0.0299 } & \multicolumn{1}{c}{0.0287 } & \multicolumn{1}{c}{0.0315 } \\
          & RosePO & \multicolumn{1}{c}{0.0732 } & \multicolumn{1}{c}{0.2885 } & \multicolumn{1}{c}{0.0322 } & \multicolumn{1}{c}{0.1817 } & \multicolumn{1}{c}{0.0101 } & \multicolumn{1}{c}{0.0389 } & \multicolumn{1}{c}{0.0251 } & \multicolumn{1}{c}{0.0266 } \\
          & SPPO  & \multicolumn{1}{c}{0.0628 } & \multicolumn{1}{c}{0.2684 } & \multicolumn{1}{c}{0.0298 } & \multicolumn{1}{c}{0.1533 } & \multicolumn{1}{c}{0.0064 } & \multicolumn{1}{c}{0.0283 } & \multicolumn{1}{c}{0.0316 } & \multicolumn{1}{c}{0.0357 } \\
         & \model{} & \textbf{0.0596(-5.1\%)} & \textbf{0.2398(-10.8\%)} & \textbf{0.0275(-7.7\%)} & \textbf{0.1289(-15.9\%)} & \textbf{0.0051(-20.3\%)} & \textbf{0.0262(-7.4\%)} & \textbf{0.0329(+4.1\%)} & \textbf{0.0373(+4.5\%)} \\
    \midrule
    \multirow{6}[1]{*}{Steam} & BIGRec & \multicolumn{1}{c}{0.0574 } & \multicolumn{1}{c}{0.1920 } & \multicolumn{1}{c}{0.0283 } & \multicolumn{1}{c}{0.3461 } & \multicolumn{1}{c}{0.0065 } & \multicolumn{1}{c}{0.0172 } & \multicolumn{1}{c}{0.0826 } & \multicolumn{1}{c}{0.0926 } \\
          & RW    & \multicolumn{1}{c}{0.0504 } & \multicolumn{1}{c}{0.1714 } & \multicolumn{1}{c}{0.0268 } & \multicolumn{1}{c}{0.3345 } & \multicolumn{1}{c}{0.0067 } & \multicolumn{1}{c}{0.0180 } & \multicolumn{1}{c}{0.0800 } & \multicolumn{1}{c}{0.0889 } \\
          & D3    & \multicolumn{1}{c}{0.0591 } & \multicolumn{1}{c}{0.2052 } & \multicolumn{1}{c}{0.0277 } & \multicolumn{1}{c}{0.3415 } & \multicolumn{1}{c}{0.0064 } & \multicolumn{1}{c}{0.0150 } & \multicolumn{1}{c}{0.0829 } & \multicolumn{1}{c}{0.0935 } \\
          & RosePO & \multicolumn{1}{c}{0.0484 } & \multicolumn{1}{c}{0.1688 } & \multicolumn{1}{c}{0.0262 } & \multicolumn{1}{c}{0.3293 } & \multicolumn{1}{c}{0.0083 } & \multicolumn{1}{c}{0.0216 } & \multicolumn{1}{c}{0.0861 } & \multicolumn{1}{c}{0.0964 } \\
          & SPPO  & \multicolumn{1}{c}{0.0410 } & \multicolumn{1}{c}{0.1596 } & \multicolumn{1}{c}{0.0247 } & \multicolumn{1}{c}{0.3041 } & \multicolumn{1}{c}{0.0058 } & \multicolumn{1}{c}{0.0145 } & \multicolumn{1}{c}{0.0927 } & \multicolumn{1}{c}{0.1043 } \\
         & \model{} & \textbf{0.0375(-8.5\%)} & \textbf{0.1541(-3.4\%)} & \textbf{0.0232(-6.1\%)} & \textbf{0.2875(-5.4\%)} & \textbf{0.0049(-15.5\%)} & \textbf{0.0128(-11.7\%)} & \textbf{0.0953(+2.8\%)} & \textbf{0.1077(+2.8\%)} \\
    \midrule
    \multirow{6}[1]{*}{ADM} & BIGRec & \multicolumn{1}{c}{0.1019 } & \multicolumn{1}{c}{0.3717 } & \multicolumn{1}{c}{0.0396 } & \multicolumn{1}{c}{0.5096 } & \multicolumn{1}{c}{0.1262 } & \multicolumn{1}{c}{0.3591 } & \multicolumn{1}{c}{0.0223 } & \multicolumn{1}{c}{0.0242 } \\
          & RW    & \multicolumn{1}{c}{0.0855 } & \multicolumn{1}{c}{0.3151 } & \multicolumn{1}{c}{0.0305 } & \multicolumn{1}{c}{0.4104 } & \multicolumn{1}{c}{0.1296 } & \multicolumn{1}{c}{0.3404 } & \multicolumn{1}{c}{0.0220 } & \multicolumn{1}{c}{0.0237 } \\
          & D3    & \multicolumn{1}{c}{0.1017 } & \multicolumn{1}{c}{0.3702 } & \multicolumn{1}{c}{0.0421 } & \multicolumn{1}{c}{0.5319 } & \multicolumn{1}{c}{0.1238 } & \multicolumn{1}{c}{0.3279 } & \multicolumn{1}{c}{0.0197 } & \multicolumn{1}{c}{0.0205 } \\
          & RosePO & \multicolumn{1}{c}{0.0826 } & \multicolumn{1}{c}{0.2967 } & \multicolumn{1}{c}{0.0288 } & \multicolumn{1}{c}{0.3885 } & \multicolumn{1}{c}{0.1207 } & \multicolumn{1}{c}{0.3211 } & \multicolumn{1}{c}{0.0212 } & \multicolumn{1}{c}{0.0226 } \\
          & SPPO  & \multicolumn{1}{c}{0.0803 } & \multicolumn{1}{c}{0.2821 } & \multicolumn{1}{c}{0.0243 } & \multicolumn{1}{c}{0.3672 } & \multicolumn{1}{c}{0.1172 } & \multicolumn{1}{c}{0.3169 } & \multicolumn{1}{c}{0.0245 } & \multicolumn{1}{c}{0.0274 } \\
         & \model{} & \textbf{0.0788(-2.0\%)} & \textbf{0.2744(-2.7\%)} & \textbf{0.0215(-11.5\%)} & \textbf{0.3347(-13.5\%)} & \textbf{0.1068(-8.9\%)} & \textbf{0.2924(-9.5\%)} & \textbf{0.0259(+5.7\%)} & \textbf{0.0294(+7.3\%)} \\
    \bottomrule
    \end{tabular}%
}
\end{table*}

\subsubsection{Hardware Information}
We run all experiments on the same Ubuntu 20.04 LTS System server with 48-core CPU, 256GB RAM, and NVIDIA A800 GPU.
%
% More experimental details are reported in Appendix ~\ref{app:exp}.
\subsection{Main Results}

\begin{figure}
    \centering
    \includegraphics[width=\linewidth]{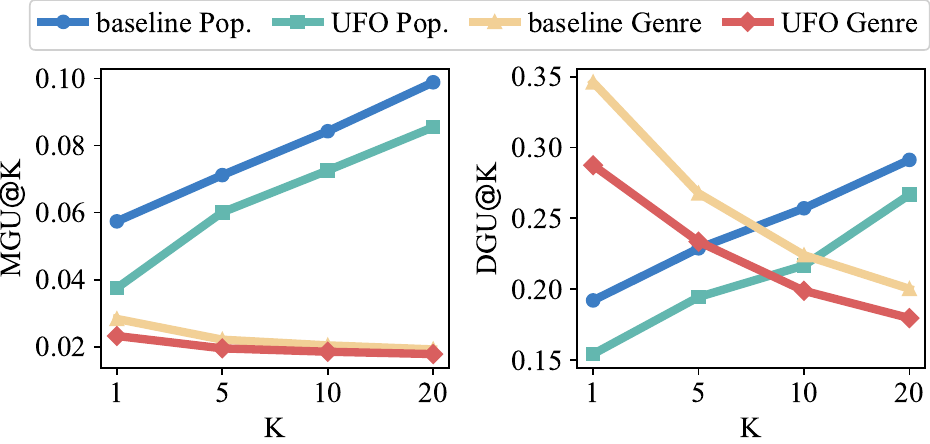}
    \caption{Fairness performance under different top-K recommendations on the Steam dataset. Pop. denotes Popularity Fairness, and Genre denotes Genre Fairness.}
    \label{fig:top-k}
    \vspace{-10pt}
\end{figure}

\paragraph{\model{} significantly improves IF in multiple dimensions} As illustrated in Table~\ref{tab:overview}, \model{} significantly enhances IF. Across three datasets and three strategies for groups split, \model{} achieves an average improvement in fairness of 9.5\% in MGU and 8.9\% in DGU. D3 does not provide a consistent improvement in IF. RW achieves moderate gains but remains limited, partly because it relies on static popularity statistics and must be retrained for each group-split strategy. RosePO and SPPO are able to continuously reduce IF, but their improvements are still insufficient. Both methods follow the DPO-style preference optimization paradigm, which lacks the ability to identify and address the underlying sources of unfairness within LRSs. In contrast, by dynamically monitoring and mitigating the inherent unfairness introduced during pre-training and SFT, \model{} achieves substantially stronger and more robust IF improvements across all datasets and fairness dimensions. Additionally, as shown in Fig.~\ref{fig:top-k}, across different top-K recommendation settings, regardless of whether unfairness increases or decreases as K grows, \model{} consistently delivers stable IF improvements.
\begin{table}[t]
\caption{Ablation results in terms of Popularity IF and recommendation performance on ML-1M. "DNG" denotes Distributional Next-item Generation, and "GM" denotes Geometric Mixture. "w/ history" denotes using history item distribution as target. "w/o" for without specific components.}
\label{tab:ablation}
\centering
% \resizebox{\textwidth}{!}{
\begin{tabular}{lcccc}
    \toprule
    \multicolumn{1}{c}{\multirow{2}[4]{*}{Method}} & \multicolumn{2}{c}{Fairness $\downarrow$} & \multicolumn{2}{c}{Accuracy $\uparrow$} \\
\cmidrule{2-5}          & \multicolumn{1}{c}{MGU} & \multicolumn{1}{c}{DGU} & \multicolumn{1}{c}{NDCG@5} & \multicolumn{1}{c}{HR@5} \\
    \midrule
    BIGRec & 0.0866 & 0.3146 & 0.0282 & 0.0310 \\
    w/o DNG & 0.0612 & 0.2513 & 0.0323 & 0.0368 \\
    w/o evolving & 0.0687 & 0.2739 & 0.0304 & 0.0331 \\
    w/o GM & 0.0622 & 0.2612 & 0.0318 & 0.0361 \\
    w/ history & \textbf{0.0558} & \textbf{0.2105} & 0.0226 & 0.0238 \\
    \model{} & 0.0596 & 0.2398 & \textbf{0.0329} & \textbf{0.0373} \\
    \bottomrule
    \end{tabular}%
% }
\vspace{-10pt}
\end{table}
\paragraph{\model{} can improve recommendation performance}
As shown in Table~\ref{tab:overview}, correctly improving fairness can also enhance recommendation quality, since reducing unfairness helps boost the performance of disadvantaged groups and push the model closer to the true underlying distribution. However, existing methods that directly modify training or decoding often fail to preserve accuracy. RW and D3 frequently degrade NDCG and HR, and RosePO also sacrifices accuracy on several datasets. SPPO is relatively more stable and can even improve accuracy, but its gains remain limited because, like RosePO, it still operates at the sample level via DPO and does not perform true distribution-level optimization.

In contrast, \model{} not only avoids harming recommendation performance but consistently achieves the best NDCG@5 and HR@5 across all datasets. On average, \model{} delivers 4.2\% and 4.9\% relative improvements in NDCG@5 and HR@5 over the strongest baseline. Because we explicitly treat the recommendation data distribution as the fair target distribution, \model{} performs genuine distribution-level alignment, which provides additional utility gains. Moreover, the introduction of geometric mixture further enhances the expressiveness and performance of the underlying LRS. Our ablation analysis further confirms \model{}’s positive influence on recommendation performance.

\begin{figure}
    \centering
    \includegraphics[width=\linewidth]{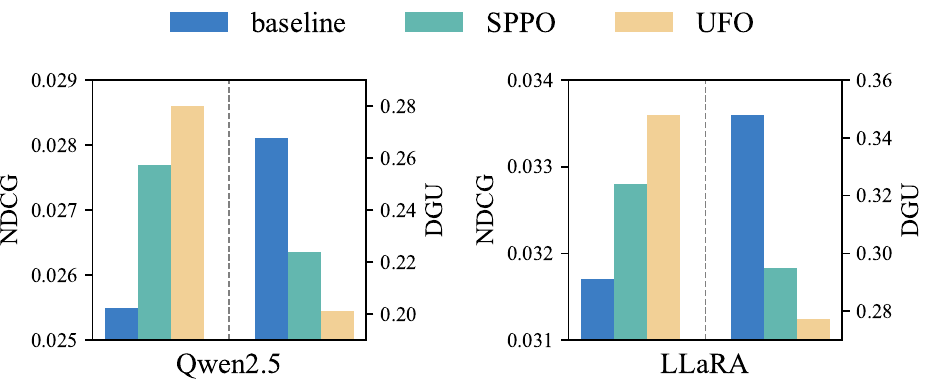}
    \caption{Results on other base LLM and LRS Model.}
    \label{fig:qwen_llara}
    \vspace{-15pt}
\end{figure}

\paragraph{\model{} is consistently effective across different LRSs and base LLMs}
As shown in Table~\ref{tab:overview} and Fig.~\ref{fig:qwen_llara}, which reports the Popularity Fairness results on the ML-1M dataset, when applied to BIGRec and LLaRA, \model{} consistently improves both fairness and recommendation quality. Likewise, across different base LLMs including Llama2 and Qwen2.5, \model{} maintains stable gains. These results highlight \model{}’s robustness and indicate that its dynamic unfairness mitigation aligns well with the intrinsic learning dynamics of modern LRSs, enabling reliable improvements regardless of the underlying LRS or base LLM.

\subsection{Ablation studies}

\begin{figure*}
    \centering
    \subfigure{
        \includegraphics[width=0.3\linewidth]{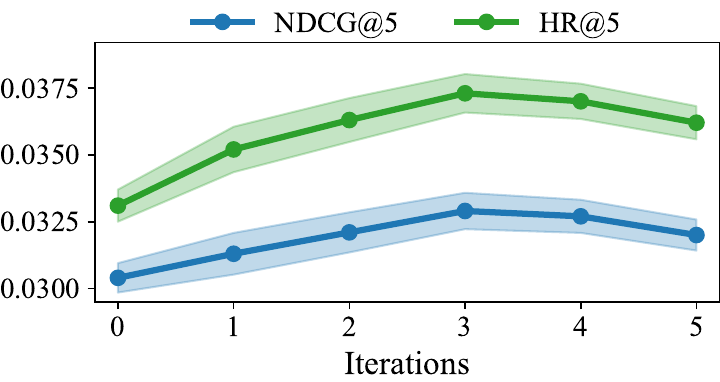}
    }
    \subfigure{
        \includegraphics[width=0.3\linewidth]{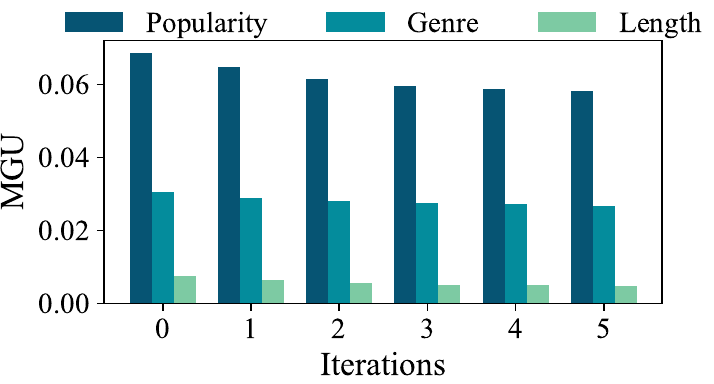}
    }
    \subfigure{
        \includegraphics[width=0.3\linewidth]{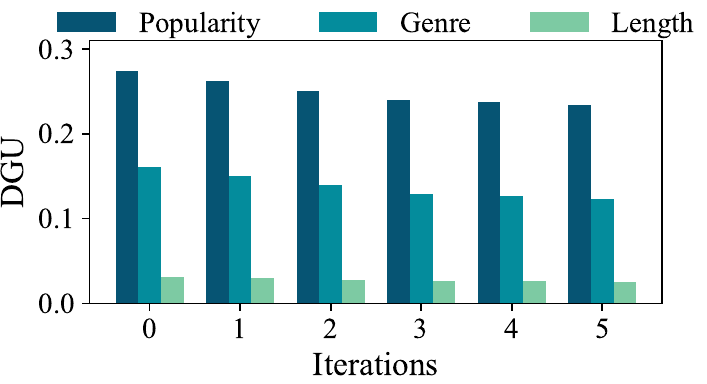}
    }
    \caption{Effect of evolving iterations.}
    \label{fig:iteration}
    \vspace{-10pt}
\end{figure*}

\begin{figure*}
    \centering
    \subfigure{
        \includegraphics[width=0.3\linewidth]{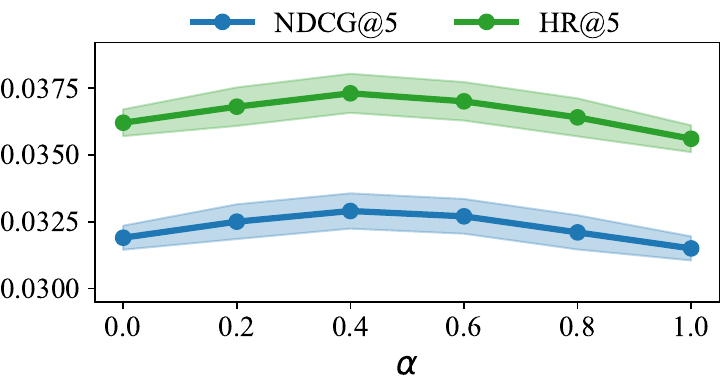}
    }
    \subfigure{
        \includegraphics[width=0.3\linewidth]{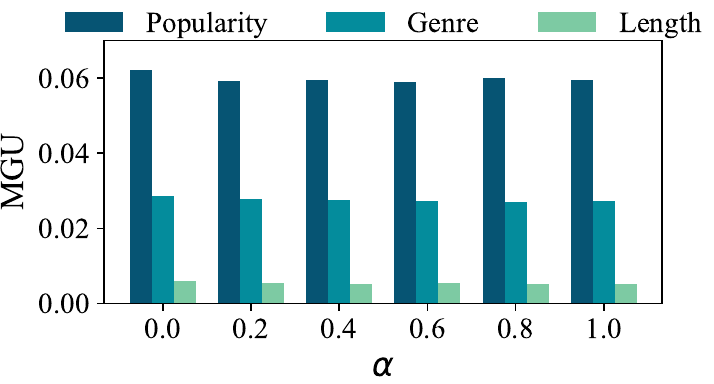}
    }
    \subfigure{
        \includegraphics[width=0.3\linewidth]{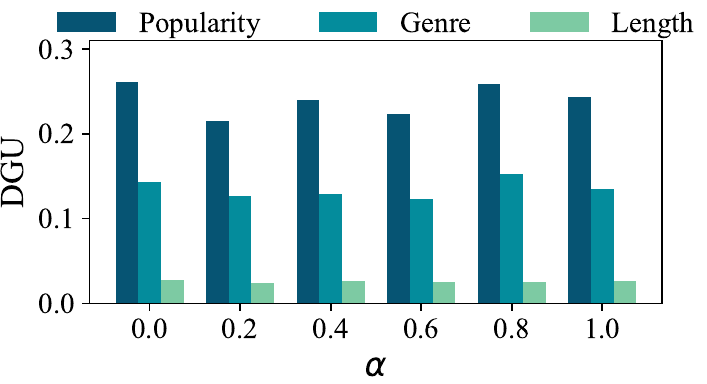}
    }
    \caption{Effect of geometric mixture parameter $\alpha$.}
    \label{fig:geometric}
    \vspace{-15pt}
\end{figure*}

\subsubsection{The effect of evolving and evolving iterations}
As shown in Table~\ref{tab:ablation} and Fig.~\ref{fig:iteration}, iteratively evolving the LRS steadily strengthens IF. The largest fairness gains appear in the early iterations, while the improvements in the fourth and fifth iterations become noticeably smaller, suggesting that the model has already corrected most prominent unfairness patterns and is gradually approaching a saturation point. Meanwhile, as the number of evolving steps increases, the contribution of geometric mixture to recommendation performance gradually weakens. The optimization shifts more toward fairness, leading to a slight drop in NDCG and HR compared with the peak iteration, although the performance remains higher than that of the 0-iteration model. This reflects the growing weight of fairness enhancement within the LRS’s effective optimization objective.

To balance IF and recommendation quality, we therefore stop the evolving process one iteration before a noticeable decline in recommendation performance is observed. In practice, the number of evolving iterations can be flexibly chosen according to application goals—for example, favoring maximal fairness or prioritizing recommendation accuracy.
\subsubsection{The effect of geometric mixture and mixture parameter $\alpha$}
We experiment with a range of \(\alpha\) values set at \([0, 0.2, 0.4, 0.6, 0.8, 1.0]\). First, when \(\alpha\) is set to 0, this condition removes the geometric mixture, causing \model{} to revert to a conventional self-play mechanism. As reported in Table~\ref{tab:ablation} and Fig.~\ref{fig:geometric}, in this configuration, \model{} indeed sacrifices a portion of recommendation performance due to the unconstrained multi-iteration evolutionary process. Conversely, setting \(\alpha\) to 1 transforms \model{} loss function to resemble that form of DPO, yet it performs suboptimally in both fairness and recommendation efficacy. This suggests that despite following the self-play training paradigm, when setting \(\alpha\) to 1, \model{} has fallen into a state of overfitting when comparing against a fixed target policy. Notably, the optimal \(\alpha\) value surpasses the extremes in both fairness and recommendation performance, indicating that the geometric mixture introduced not only benefits recommendation performance but also exhibits superior fairness compared to standalone regularizer.

\subsubsection{The effect of distributional next-item generation}
As shown in Table~\ref{tab:ablation}, removing the distributional next-item generation module leads to clear degradation in both IF and recommendation performance. Without this component, \model{} loses its ability to operate at the distribution level, resulting in evolution that is no longer guided toward the underlying fair distribution.

\subsubsection{The choice of target fair distribution}
Since the definition of IF uses the historically interacted items as the reference, we further consider another target distribution—the history item distribution, where target samples are drawn according to historical proportions. As shown in Table~\ref{tab:ablation}, this choice “cheats” the IF metric and thus yields higher IF, but it leads to a severe drop in recommendation performance. Moreover, the improvement in IF remains limited because the history distribution differs substantially from the SFT training objective, making it difficult for the LRS to effectively evolve toward this new target. This also highlights the superiority of using the original recommendation data distribution as the target fair distribution.
\section{Conclusion}
In this paper, we systematically analyze the joint impact of pre-training and SFT on LRSs fairness, providing an in-depth explanation of the root causes of severe unfairness in LRSs.
Based on this finding, we propose \model{} method as a post-training solution to address unfairness. \model{} directly addresses all forms of unfairness in LRSs, while maintaining recommendation performance. 
During the iterative evolution process, different iterations of LRS alternately play as the judger (fine-tuned to better identify unfairness) and the corrector (correcting unfairness so generating more fair recommendation results), enabling fairness enhancement through a self-play mechanism.
Additionally, we incorporate a geometric mixture policy, leveraging the geometric mixture between the original LRS and the LRS from the last iteration as a constraint to preserve the recommendation performance. 
Extensive experimental results demonstrate that \model{} significantly improves the fairness of LRSs while achieving high-level recommendation performance.

%% The file named.bst is a bibliography style file for BibTeX 0.99c
\bibliographystyle{ijcai/named}
\bibliography{ref}

\end{document}